\input phyzzx

\def\tr{\rm tr}
%%%%%%%If you do not have the msbm fonts, delete the following 4 lines

\def\cF {{\cal F}}
%%%%%%%%%%%%

%%%%%%%%%%%%%%%%%%%%%%%%%%%%%%%%%%%%%%%%%%%%%%%%%%%%%%%%%%%%%%%%%%%%%%%%%%%%%
\REF\PKTa{P.K. Townsend, Phys. Lett. {\bf B373} (1996) 68.}
\REF\HS{P.S. Howe and E. Sezgin, Phys. Lett. {\bf 390B} (1997) 133.}
\REF\swedes{M. Cederwall, A von Gussich, B.E.W. Nilsson and A. Westerberg,
Nucl. Phys. {\bf B490} (1997) 165}
\REF\swedesb{M. Cederwall, A von Gussich, B.E.W. Nilsson,
P. Sundell and A. Westerberg, Nucl. Phys. {\bf B490} (1997) 179.}
\REF\bergstown{E. Bergshoeff and P.K. Townsend,
Nucl. Phys. {\bf B490} (1997) 145.}
\REF\JHS{M. Aganagic, C. Popescu and J.H. Schwarz, {\sl D-brane actions with
local $\kappa$-symmetry}, hep-th/9610249; {\sl Gauge-invariant and gauge-fixed
D-brane actions}, hep-th/9612080.}
\REF\bandos{I. Bandos, D.P. Sorokin and M. Tonin, {\sl Generalized action
principle and superfield equations of motion for D=10 D-p-branes},
hep-th/9701127.}
\REF\wita{E. Witten, Nucl. Phys. {\bf B266} (1986) 245.}
\REF\gris{M.T. Grisaru, P.S. Howe, L. Mezincescu, B.E.W. Nilsson and P.K.
Townsend, Phys. Lett. {\bf 162B} (1985) 116; E. Bergshoeff, E. Sezgin and P.K.
Townsend, Phys. Lett. {\bf 169B} (1986) 191.}
\REF\BST{E. Bergshoeff, E. Sezgin and P.K. Townsend, Phys. Lett. {\bf  189B}
(1987) 75; Ann. Phys. (N.Y.) {\bf 185} (1988) 330.}
\REF\DHIS{M.J. Duff, P.S. Howe. T. Inami and K.S. Stelle, Phys. Lett.
{\bf B191}, (1987) 70.}
\REF\romans{L. Romans, Phys. Lett. {\bf B169} (1986) 374.}
\REF\bdr{E. Bergshoeff and M. de Roo, Phys. Lett. {\bf 380B} (1996) 265.}
\REF\ght{M.B. Green, C.M. Hull and P.K. Townsend, Phys. Lett. {\bf 382B} (1996)
65.}
\REF\topmass{W. Siegel, Nucl. Phys. {\bf 156B} (1979) 135; J. Schonfeld, Nucl.
Phys. {\bf 185B} (1981) 135; R. Jackiw and S. Templeton, Phys. Rev. {\bf D23}
(1981) 2291; S. Deser, R. Jackiw and S. Templeton, Phys. Rev. Lett. {\bf 48}
(1982) 975; Ann. Phys. (N.Y.) {\bf 140} (1982) 372.}
\REF\eightbrane{E. Bergshoeff, M. de Roo, M.B. Green, G. Papadopoulos and P.K.
Townsend, Nucl. Phys. {\bf B470} (1996) 113.}
\REF\chpt{P.M. Cowdall, H. L{\" u}, C.N. Pope, K.S. Stelle and P.K. Townsend,
Nucl. Phys. {\bf B486} (1997) 49.}
\REF\gates{J.L. Carr, S.J. Gates, Jr., and R.N. Oerter, Phys. Lett. {\bf189}
(1987) 68.}
\REF\BHO{E. Bergshoeff, C.M. Hull and T. Ort\'\i n, Nucl. Phys. {\bf B451}
(1995) 547.}
\REF\berone{E. Bergshoeff, M. de Roo and T. Ort\'\i n, 
Phys. Lett. {\bf B386} (1996) 85.}
\REF\brink{E. Cremmer and S. Ferrara, Phys. Lett. {\bf 91B} (1980) 61; L. Brink
and P.S. Howe, Phys. Lett. {\bf 91B} (1980) 153.}
\REF\bertwo{ E. Bergshoeff, {\sl p-branes, D-branes and M-branes}, hep-th/9611099}

%%%%%%%%%%%%%%%%%%%%%%%%%%%%%%%%%%%%%%%%%%%%%%%%%%%%%%%%%%%%%%%%%%%%

\Pubnum{ \vbox{ \hbox{R/97/20} \hbox{UG-6/97} \hbox{hep-th/9706094}} }
\pubtype{}
\date{June 1997}

\titlepage

\title {\bf Massive IIA supergravity from the topologically massive D-2-brane}

\author{E. Bergshoeff}
\address{Institute for Theoretical Physics, Nijenborgh 4,
\break
9747 AG Groningen, The Netherlands}
\andauthor{P.M. Cowdall and P.K. Townsend}
\address{DAMTP, University of Cambridge,
\break
Silver St., Cambridge CB3 9EW, U.K.}

\abstract{The superfield equations of massive IIA supergravity, in the form of
constraints on the superspace geometry, are shown to be implied by
$\kappa$-symmetry of the topologically massive D-2-brane.}

\endpage
%\pagenumber=1

\chapter{Introduction}

The problem of the determination of the full $\kappa$-symmetric action for type
II super D-branes in general supergravity backgrounds has now been largely
solved by the concerted efforts of several groups
[\PKTa,\HS,\swedes,\swedesb,\bergstown,\JHS,\bandos]. Most of this work has
concentrated on the verification of $\kappa$-symmetry in backgrounds of varying
generality, but it is known from earlier work on super p-branes
[\wita,\gris,\BST,\DHIS] that the requirement of $\kappa$-symmetry constrains
the possible backgrounds. For example, $\kappa$-symmetry of the D=11
supermembrane requires the background to satisfy the field equations of D=11
supergravity. Moreover, since
$\kappa$-symmetry is necessary for the consistency of the worldvolume field
equations, i.e. the `branewave' equations, one can view the equations of 
D=11 supergravity as branewave integrability conditions. The field
equations of D=10 IIA supergravity similarly follow from $\kappa$-symmetry of 
the IIA D-2-brane; we shall verify this here, but it is an immediate
consequence of the fact that the super D-2-brane action is dual to the
D=11 supermembrane action in D=11 backgrounds with a $U(1)$ isometry
[\bergstown]. Such a background is equivalent to one of D=10 IIA supergravity.
However, not all IIA backgrounds can be viewed as reductions of D=11
supergravity backgrounds. Specifically, only the usual, `massless', IIA
supergravity is obtainable in this way. The `massive' IIA theory, which has a
cosmological constant proportional to a mass parameter $m$ [\romans], has no
known interpretation of this type, although one might expect the field equations
to be required by $\kappa$-symmetry of some generalization of the super
D-2-brane action. In fact, it was  shown in [\bergstown] that $\kappa$-symmetry 
of the D-2-brane action in a purely bosonic IIA background requires the inclusion
of a worldvolume Chern-Simons term when $m\ne0$, as expected from earlier
T-duality considerations [\bdr,\ght]. We shall refer to this as the
`topologically massive' D-2-brane, since the CS term constitutes a
topological mass term for the Born-Infeld field [\topmass]. 

The main purpose of this letter is to show that the massive IIA
field equations are a {\sl consequence} of $\kappa$-symmetry of the 
topologically massive D-2-brane action; similar results
then follow for lower-dimensional massive supergravity theories
[\eightbrane,\chpt] by dimensional reduction. To establish this requires
consideration of general IIA supergravity backgrounds, including fermions. It is
notable that the massive field equations obtained in this way arise as a
particularly simple set of superfield constraints that are formally the same as
those of the massless IIA theory, differing only in the $m$-dependence of the
field strengths. Superspace constraints for both massless and massive IIA
supergravity have been proposed previously [\gates]. It is not clear to us
whether our results are in complete agreement with those of [\gates]. In any
case, we think it worthwhile to have an independent derivation of these
constraints in view of the fact that invariance under supersymmetry was not
completely established in [\romans] because terms quartic in fermions were
omitted from the action.

The coupling of D-branes to a supergravity background leads to a particular
basis of supergravity field variables. As seen in [\ght], and as we shall see
again here, this basis leads to a number of simplifications as compared to the
`canonical' basis used in the supergravity literature (e.g. [\BHO,\berone]). 
We conclude this paper with an examination of the details of the map from the 
old variables to the new `D-brane inspired' ones.
 
%%%%%%%%%%%%%%%%%%%%%%%%%%%%%%%%%%%%%%%%%%

\chapter{The D-2-brane in a general IIA background}

Let $Z^M$ be local coordinates on D=10 IIA superspace, with $E_M{}^A$ the
superspace vielbein, so $E^A\equiv dZ^M E_M{}^A$ is a basis of one-forms on
superspace. We define a worldvolume metric, in local coordinates
$\xi^i$, by
$$
g_{ij}= E_i{}^a E_j{}^b\,\eta_{ab}\ ,
\eqn\oneaa
$$ 
where $E_i{}^A= \partial_i Z^ME_M{}^A$ and
$\eta$ is the D=10 Minkowski metric. We introduce a scalar superfield
$\phi$ and two-form superspace gauge potential $B$, the lowest components of
which are, respectively, the dilaton and the Neveu-Schwarz/Neveu-Schwarz
($NS\otimes NS$) two-form gauge potential. We also introduce a Born-Infeld
1-form gauge potential $V$ with `modified' field strength 
$$
\cF = dV -B\, .
\eqn\onea
$$
Whereas $V$ is defined directly on the worldvolume the two form $B$ is here the
pullback of the two-form on superspace; we use the same letter to denote the
superspace and worldvolume forms since it should be clear from the context
which is intended. Finally, we introduce the superspace 1-form $C$ and 3-form
$A$, the lowest components of which are the Ramond/Ramond ($R\otimes R$) gauge
potentials. Again, we shall use the same letters to denote their pullbacks to
the worldvolume. With these ingredients we can write down the action for the
super D-2-brane in a general IIA supergravity background. Setting the
tension to unity we have
$$
S= -\int d^3\xi\, \bigg[ e^{-\phi}\sqrt{-\det(g+ \cF )} \, +\,
{1\over6}\varepsilon^{ijk} (A_{ijk} + 3C_i\cF_{jk} + {3\over2} 
mV_iF_{jk})\bigg]\, ,
\eqn\oneb
$$
where $m$ is the mass parameter. 

The structure group of the superspace tangent bundle is taken to be the Lorentz
group, with respect to which $E^A$ decomposes into $E^A=(E^a,E^\alpha)$ where
$E^a$ is a Lorentz vector and $E^\alpha$ a Majorana spinor. The spacetime Dirac
matrices $\Gamma_a$ can be pulled back to the worldvolume to yield
$$
\gamma_i = E_i^a \Gamma_a\ ,
\eqn\oned
$$
which behave like three-dimensional Dirac matrices except for the fact that the
product of all three is not the identity matrix. Instead, the matrix
$$
\Gamma_{(0)} = {1\over6\sqrt{-\det g}}\varepsilon^{ijk} \gamma_{ijk}
\eqn\onee
$$
is traceless, commutes with $\gamma_i$ matrices, and satisfies
$$
\Gamma_{(0)}^2 =1 \ .
\eqn\onef
$$
We refer to [\bergstown] for further details of the
notation and conventions, but we note here that the exterior derivative of a
scalar superfield $\phi$ can be expanded on the basis of 1-forms $E^A=dZ^M
E_M{}^A$ as
$$
d\phi = E^A D_A\phi\ ,
\eqn\onec
$$
which defines the supercovariant derivative $D_A\phi$ of $\phi$. 

To present the $\kappa$-symmetry variations it will be convenient to define 
$\delta E^A := \delta Z^M E_M{}^A$. The variation of the worldvolume fields $Z^M$
is always such that $\delta_\kappa E^a=0$. Making use of various lemmas
presented in [\bergstown] we then find that
$$
\eqalign{
\delta_\kappa \phi &= \delta_\kappa E^\alpha D_\alpha \phi \cr
\delta_\kappa g_{ij} &= -2\delta_\kappa E^\alpha E_{(i}{}^a E_{j)}{}^A
T_{A\alpha}{}^b
\eta_{ab}\cr
\delta_\kappa C_i &= \partial_i(\delta_\kappa E^\alpha C_\alpha) -\delta_\kappa
E^\alpha E_i{}^B (K-mB)_{B\alpha} \cr
\delta_\kappa A_{ijk} &= \delta_\kappa E^\alpha E_i{}^B E_j{}^C E_k{}^D
\big\{F_{DCB\alpha} -4C_{[D}H_{CB\alpha]} -3m B_{DC}B_{B\alpha}\big\} \cr
&\qquad\qquad + {\rm total\, derivative} }
\eqn\oneh
$$
where $T_{AB}{}^C$ is the superspace torsion, $H=dB$ is the $NS\otimes NS$
two-form field strength, and 
$$
\eqalign{
K &= dC + mB\cr
F &= dA + H\wedge C + {1\over2} m B\wedge B}
\eqn\oneha
$$
are the $R\otimes R$ superspace field strengths [\ght,\bergstown]. The square
brackets around suffices indicate super-antisymmetrization on the enclosed
indices. Note that we adopt the conventions
$$
\eqalign{
P &= {1\over p!} E^{A_p}\dots E^{A_1} P_{A_1\dots A_p}\cr
d(PQ) &= PdQ + (-1)^q (dP)Q }
\eqn\conven
$$
for superspace $p$-form $P$ and $q$-form $Q$, where the exterior product of
forms is understood. These conventions lead to some sign differences relative to [\ght]. We adopt the same conventions for worldvolume forms, e.g.
$$
{\cal F} = {1\over2}d\xi^j d\xi^i\, {\cal F}_{ij}\, ,
\eqn\convenb
$$
which leads to some sign differences relative to [\bergstown].

Apart from the specification of $\delta_\kappa E^\alpha$, which will be 
postponed until later, we must also specify $\delta_\kappa V$. This must be such
as to ensure that the variation of $\cF$ is `supercovariant', i.e. appears
without derivatives of the parameter $\kappa$, and this property essentially
fixes it uniquely to be
$$
\delta V_i = E_i{}^A\delta_\kappa E^B B_{BA}\ .
\eqn\onehab
$$
The resulting transformation of $\cF$ is\foot{The sign differs from 
[\bergstown] as a result of the `reverse order' convention \convenb\ for
the components of worldvolume forms.}
$$
\delta_\kappa \cF_{ij} = E_i{}^A E_j{}^B\delta_\kappa E^\alpha 
H_{\alpha BA}\ .
\eqn\onei
$$
With these variations in hand, and discarding a surface term, we
compute that
$$
\eqalign{
\delta_\kappa S &= \int d^3\xi\, \delta_\kappa E^\alpha \bigg\{
e^{-\phi}\sqrt{-\det(g+ \cF )}
\big[ D_\alpha\phi \cr
& \qquad +(g+\cF)^{ij}\big( E_{(i}{}^a
E_{j)}{}^B T_{B\alpha}{}^c\eta_{ac} + {1\over2}E_i^A E_j{}^B
H_{BA\alpha}\big)\big] \cr
& +{1\over6}\varepsilon^{ijk}\big( E_k{}^A E_j{}^B E_i{}^C F_{CBA\alpha} +
3E_i^A\cF_{jk} K_{A\alpha} \big) \bigg\}\, , }
\eqn\onej
$$
where $(g+\cF)^{ij}$ are the entries of the inverse of the matrix $(g+\cF)$.
Note that all $m$-dependence of this variation is now implicit in the $R\otimes
R$ field strengths. 

Following [\bergstown], it is convenient to introduce the matrix $X$ by
$X=g^{-1}{\cal F}$, or
$$
X^i{}_j = g^{ik}{\cal F}_{kj}
\eqn\extraa
$$
Because of the antisymmetry of ${\cal F}$, this matrix satisfies the identity
$$
X^3 \equiv {1\over2} (\tr X^2) X \, .
\eqn\extrab
$$
We now rewrite \onej\ as a sum of terms in which each term involves a different
number of worldvolume fermions. Thus,
$$
\delta_\kappa S = \int d^3\xi\, \sqrt{-\det g}\, e^{-\phi}\, 
\delta_\kappa E^\alpha\, [ \Delta_0 + \Delta_1 + \Delta_2 + \Delta_3]_\alpha
\eqn\oneja
$$
where
$$
\eqalign{
(\Delta_0)_\alpha  &=  \sqrt{\det (1+X)}\,\big\{[(1+X)^{-1}]^i{}_k\, g^{kj}
\big [E_{(i}{}^a E_{j)}{}^b T_{b\alpha}{}^c\eta_{ca} +{1\over2} E_i{}^a
E_j{}^b H_{ba\alpha}\big] \cr
&  + D_\alpha\phi\big\}   + {e^\phi\over6\sqrt{-\det g}} 
\varepsilon^{ijk} [E_k{}^a E_j{}^b E_i{}^c F_{cba\alpha} + 3E_i{}^a (gX)_{jk}
K_{a\alpha}] \cr 
(\Delta_1)_\alpha  &= \sqrt{\det (1+X)}\, 
[(1+X)^{-1}]^i{}_k\, g^{kj}\big[E_{(i}{}^a E_{j)}{}^\beta
T_{\beta\alpha}{}^c\eta_{ca} - E_{[i}{}^a E_{j]}{}^\beta H_{a\beta\alpha}\big]
\cr  &\qquad + {e^\phi\over2\sqrt{-\det g}} \varepsilon^{ijk} E_i{}^\beta
\big[ E_j{}^a E_k{}^b F_{ab\beta\alpha} + (gX)_{jk}K_{\beta\alpha}\big] \cr
(\Delta_2)_\alpha  &= {1\over2} E_i{}^\beta E_j{}^\gamma\,\big\{
\sqrt{\det (1+X)}\, [(1+X)^{-1}]^i{}_k\, g^{kj}H_{\gamma\beta\alpha} \cr
&\qquad - {e^{\phi}\over\sqrt{-\det g}} \varepsilon^{ijk}E_k{}^c
F_{c\beta\gamma\alpha} \big\}\cr 
(\Delta_3)_\alpha  &= {e^\phi\over6\sqrt{-\det g}} \varepsilon^{ijk} 
E_k{}^\beta E_j{}^\gamma E_i{}^\delta F_{\delta\gamma\beta\alpha} \ .}
\eqn\onejb
$$
The subscript on $\Delta$ indicates the number of worldvolume fermions.
Each of these terms must vanish separately, for some choice of $\delta_\kappa
E^\alpha$. 

We know that $\delta_\kappa E^\alpha$ must take the form
$$
\delta_\kappa E^\alpha = [\bar\kappa (1-\hat\Gamma)]^\alpha\ ,
\eqn\forma
$$
where the matrix $\hat\Gamma$ is tracefree and squares to the identity matrix. 
From [\bergstown] we know that for backgrounds that are purely bosonic
solutions of IIA supergravity we can choose\foot{We refer to [\bergstown] for 
the relation of $\hat\Gamma$ to the `standard' matrix $\Gamma$ that arises in 
the proof of $\kappa$-invariance for general $p$.} 
$$
\hat\Gamma = \sqrt{(1-{1\over2}\tr X^2)}\, \Gamma_{(0)} + {1\over2}
X_{ij}\Gamma^{ij} \Gamma_{11}\, .
\eqn\onekb
$$
We do not wish to assume here that this is the form of $\hat\Gamma$ in general
backgrounds, although this will turn out to be the case. We shall need only
the expansion to first order in $X$, which is
$$
\hat\Gamma = \Gamma_{(0)} -{1\over2} \gamma^{ij}X_{ij}\Gamma_{11} + 
{\cal O}(X^2)\, .
\eqn\formb
$$
The leading term, independent of $X$ is one of only two possibilities
consistent with spacetime and worldvolume Lorentz invariance; the other
possibility is $\Gamma_{11}$ but this choice leads immediately to much stronger
constraints on the background so we may discard it. The term linear in $X$ is
also effectively unique; there is a freedom to replace $\Gamma_{11}$ by
$\Gamma_{(0)}\Gamma_{11}$ since this affects only the ${\cal O}(X^2)$ terms,
but this leads to equivalent results [\bergstown].  

We now turn to an examination of each of the four $\Delta$ terms in \oneja.
The term involving $\Delta_3$ can cancel only if 
$$
F_{\alpha\beta\gamma\delta}=0\ .
\eqn\twoa
$$ 
Consideration of the terms independent of and linear in $X$ in the $\Delta_2$
term leads directly to the conclusion that 
$$
H_{\alpha\beta\gamma}=0 \, ,\qquad F_{\alpha\beta\gamma\, a}=0 \ .
\eqn\twob
$$
We turn next to $\Delta_1$. The vanishing of $\delta_\kappa E\Delta_1$ to zeroth
order in $X$ requires
$$
(1-\Gamma_{(0)})^{\gamma\beta}\big[\sqrt{-\det g}\, g^{kj} E_j{}^a 
T_{\beta\alpha}{}^c\eta_{ca} + {1\over2}
\varepsilon^{ijk} \, E_i{}^a E_j{}^b e^\phi F_{ab\beta\alpha} \big] =0\, .
\eqn\twoc
$$
Without the $(1-\Gamma_{(0)})$ factor, terms with different numbers of $E_i{}^a$
factors would have to cancel separately. This would impose very strong
constraints on the background. In fact, the constraints are weaker because the
identity
$$
(1 - \Gamma_{(0)})\big[ \sqrt{-\det g}\,\gamma^i +
{1\over2}\varepsilon^{ijk}\gamma_{jk}\big] \equiv 0
\eqn\twod
$$
allows a cancellation between terms, but this can happen only if\foot{The
factor of $i$ is needed for reality of $\delta_\kappa S$ with standard
conventions for complex conjugation of products of anticommuting spinors.}
$$
T_{\beta\alpha}{}^c= i (\Gamma^c)_{\alpha\beta}\, , \qquad 
F_{ab\beta\alpha} = i e^{-\phi}\,(\Gamma_{ab})_{\alpha\beta}\, .
\eqn\twoe
$$
In principle, these expressions could come multiplied by some scalar function
but this could be removed by a rescaling of the component of the spin connection
that these `conventional' constraints allow us to solve for. We may now use
\twoe\ in the terms linear in $X$ in the expansion of
$\delta_\kappa E\Delta_1$ to find 
$$
\eqalign{
0= {1\over2}(gX)_{ij}&[(1-\Gamma_{(0)})\gamma^{ijk}]^{\gamma\beta} [e^{\phi}
K_{\beta\alpha} - (\Gamma_{11})_{\beta\alpha}] \cr
&\qquad + X^{ik}(1-\Gamma_{(0)})^{\gamma\beta}[ E_i{}^a H_{a\beta\alpha} -
(\gamma_i\Gamma_{11})_{\beta\alpha} ]\, , }
\eqn\twof
$$
from which we deduce that
$$
H_{a\alpha\beta} = i(\Gamma_a\Gamma_{11})_{\alpha\beta} \, ,
\qquad K_{\alpha\beta} = i e^{-\phi} (\Gamma_{11})_{\alpha\beta} \, .
\eqn\twog
$$
We have now determined that the superspace constraints \twoe\ and \twog\ are
necessary for $\kappa$-symmetry. Since the $\Delta_1$ terms involve no
background fermions it follows from the results of [\swedesb,\bergstown] that 
the these constraints are also sufficient for the cancellation of the $\Delta_1$
terms to all orders in $X$ if $\hat\Gamma$ is given by \onekb. 

We now turn to the $\Delta_0$ terms in \oneja, which involve
background fermion fields. We first expand expand $\delta_\kappa E\Delta_0$ in
powers of $X$. To zeroth order we find that
$$
(1- \Gamma_{(0)}) \big[ \lambda + g^{ij}E_i{}^a E_j{}^b
T_{b\alpha}{}^c\eta_{ca}
+{e^\phi\over6\sqrt{-\det g}} \varepsilon^{ijk} 
E_k{}^a E_j{}^b E_i{}^c F_{cba\alpha} \big] =0 \, ,
\eqn\extrac
$$ 
where we have introduced the dilatino superfield
$$
\lambda_\alpha = D_\alpha\phi\, .
\eqn\extrad
$$
As before, terms with different numbers of $E_i{}^a$ factors would have to
cancel separately were it not for the possibility of combining them by means of
the identity \twod\ and the further identity $E_i\cdot E_j \equiv g_{ij}$. We 
thereby deduce that the vanishing of the $\Delta_0$ terms requires
$$
T_{c\alpha}{}^b = \delta_c^b\, \chi_\alpha\ , \qquad 
F_{abc\gamma} = e^{-\phi} [\Gamma_{abc}(\lambda +3\chi)]_\gamma\ .
\eqn\threea
$$
where $\chi$ is some background spinor field. We may choose $\chi$ at will
since the torsion constraint defining $\chi$ is a `conventional' one that just
determines some components of the spin connection. Obvious choices are
$\chi=0$ and $\chi=-3\lambda$, but neither of these turns out to be the 
simplest one so we leave $\chi$ free at present.

If we now use \threea\ in the terms in $\delta_\kappa E\Delta_0$ linear in $X$
we find that
$$
\eqalign{
0 = &\sqrt{-\det g}\,
(gX)_{ij}\big[(1-\Gamma_{(0)})\gamma^{ij}\Gamma_{11}(\lambda + 3\chi)\big]^\beta
\cr & - \sqrt{-\det g}\, X^i{}_k g^{kj}\, E_i{}^a E_j{}^b
(1-\Gamma_{(0)})^{\beta\alpha} H_{ba\alpha} \cr
&\qquad +  e^\phi \varepsilon^{ijk} E_i{}^a (gX)_{jk}
(1-\Gamma_{(0)})^{\beta\alpha} K_{a\alpha}\, . }
\eqn\threeb
$$
It follows by a reasoning similar to that used previously that
$$
H_{ab\gamma} = [\Gamma_{ab}\zeta]_\gamma\ , \qquad 
K_{a\beta} = e^{-\phi}[\Gamma_a\xi]_\beta \ ,
\eqn\threec
$$
where $\zeta$ and $\xi$ are two further spinor fields. If this information is
now used in \threeb\ one finds that
$$
\xi + \zeta = -\Gamma_{11}(\lambda + 3\chi)\, .
\eqn\threed
$$
At this point we have found the general form of the constraints in terms of the
dilatino superfield and two other undetermined spinor superfields, one
combination of which must be fixed by the cancellation of terms higher order
in $X$ in the $kappa$-symmetry variation (for consistency with known results for
D=11 supermembrane. Indeed, with $\hat\Gamma$ given by
\onekb\ one finds that the relation
$$
\zeta= -2\Gamma_{11}\chi
\eqn\threee
$$
is needed for cancellation of terms quadratic in $X$, and that all higher order
terms then cancel. Thus
$$
H_{ab\gamma} = -2[\Gamma_{ab}\Gamma_{11}\chi]_\gamma \qquad 
K_{a\beta} = -e^{-\phi}[\Gamma_a\Gamma_{11}(\lambda + \chi)]_\beta \, .
\eqn\threec
$$
We now see that there is another obvious choice for $\chi$, namely
$$
\chi =-\lambda\, 
\eqn\threed
$$
since $K_{a\alpha}$ then vanishes. This choice greatly simplifies the analysis
of the Bianchi identities, to which we now turn.

%%%%%%%%%%%%%%%%%%%%%%%%%%%%%%%%%%%%%%%

\chapter{Bianchi identities}

The superspace constraints derived above are all $m$-independent. The
$m$-dependence is implicit in the $R\otimes R$ field strengths $K$ and $F$, defined in \oneha, which results in an $m$-dependence of the Bianchi
identities. These are
$$
\eqalign{
dT^A &= E^B R_B{}^A \cr
dH &=0 \cr
dF &= H\wedge K \cr
dK &= mH \ ,}
\eqn\bianchi
$$
where $R_B{}^A$ is the curvature 2-form. At dimension zero or less the Bianchi
identities are indeed satisfied by superspace tensors satisfying the 
constraints found above. In particular, the $F$ Bianchi identity at dimension
zero is satisfied by virtue of the gamma-matrix identity
$$
(\Gamma^a)_{(\alpha\beta}(\Gamma_{ab})_{\gamma\delta)} +
(\Gamma_{11})_{(\alpha\beta}(\Gamma_{11}\Gamma_b)_{\gamma\delta)}\equiv 0
\eqn\gamiden
$$
which is clearly the dimensional reduction to D=10 of the D=11 identity
required for $\kappa$-symmetry of the D=11 supermembrane [\BST].

Because the structure group of the frame bundle is taken to be the Lorentz group,
the Bianchi identities determine the only remaining torsion component at
dimension $1/2$, $T_{\alpha\beta}{}^\gamma$. The result given in [\swedesb],
where the choice
$\chi=0$ was made, is rather complicated. Here we shall see that the choice
$\chi=-\lambda$ leads to considerable simplifications. With this choice, the
Bianchi identity for $K$ at dimension 1/2 (which is $m$-independent since
$H_{\alpha\beta\gamma}=0$) implies that
$$
(\Gamma_{11})_{\epsilon(\gamma} T_{\alpha\beta)}{}^\epsilon
=(\Gamma_{11})_{(\alpha\beta}\lambda_{\gamma)}\, ,
\eqn\biana
$$
while the torsion Bianchi identity at dimension 1/2 implies that
$$
(\Gamma^a)_{\epsilon(\gamma} T_{\alpha\beta)}{}^\epsilon
=(\Gamma^a)_{(\alpha\beta}\lambda_{\gamma)}\, .
\eqn\bianb
$$
These are solved by
$$
T_{\alpha\beta}{}^\gamma = \delta_{(\alpha}^\gamma \lambda_{\beta)}\ .
\eqn\bianc
$$

We have now arrived at a set of constraints on all superspace tensors of
dimension 1/2 or less in terms of the dilatino superfield $\phi$ (since
$\lambda=D\phi$). These constraints are as follows, in order of increasing
dimension. At dimension $-1$:
$$
F_{\alpha\beta\gamma\delta} =0 \ .
\eqn\constrainta
$$
At dimension $-1/2$:
$$
H_{\alpha\beta\gamma} =0\ , \qquad  \qquad
F_{\alpha\beta\gamma\, a} =0\ .
\eqn\constraintb
$$
At dimension $0$:
$$
\eqalign{
T_{\beta\alpha}{}^a = i(\Gamma^a)_{\beta\alpha}\ ,\qquad & \qquad 
H_{\alpha\beta c} = i(\Gamma_c\Gamma_{11})_{\alpha\beta} \ ,\cr
K_{\beta\alpha} =  ie^{-\phi} (\Gamma_{11})_{\beta\alpha}\ ,\qquad &\qquad 
F_{\alpha\beta ba} = ie^{-\phi} (\Gamma_{ba})_{\alpha\beta}\ . }
\eqn\constraintc
$$
At dimension $1/2$:
$$
\eqalign{
T_{a\beta}{}^c = -\delta_a^c\lambda_\beta \ , \qquad & \qquad
T_{\beta\alpha}{}^\gamma = \delta_{(\beta}^\gamma \lambda_{\alpha)}\cr
 H_{\alpha bc} = 2(\Gamma_{bc}\Gamma_{11}\lambda)_\alpha\ , \qquad &\qquad 
 K_{a\beta} = 0 \cr
F_{\alpha abc} = -2e^{-\phi}[\Gamma_{abc} \lambda ]_\alpha \ \qquad & }
\eqn\constraintd
$$

The only undetermined components of the torsion and field strengths are now 
those of dimension $1$ or higher. These include the bosonic field strengths
$K_{ab}$,
$F_{abcd}$ and $H_{abc}$ and the torsion component $T_{\alpha b}{}^\gamma$ at
dimension $1$. These will be $m$-dependent, in general, because of the
$m$-dependence of the Bianchi identity for $K$. For example, the Bianchi 
identity for $K$ at dimension 1 is
$$
(D_a\phi) (\Gamma_{11})_{\beta\gamma} + 2 T_{a(\beta}{}^\alpha
(\Gamma_{11})_{\beta)\alpha} - e^\phi\, (\Gamma^b)_{\beta\gamma} K_{ba} +
me^\phi\, (\Gamma_a\Gamma_{11})_{\beta\gamma}=0\, .
\eqn\bianf
$$
This implies that
$$
T_{a\beta}{}^\gamma = \bar T_{a\beta}{}^\gamma - {1\over2}m e^\phi 
(\Gamma_a)^\gamma{}_{\beta}
\eqn\biang
$$
where $\bar T^A$ is the torsion 2-form for $m=0$. This $m$-dependent
modification of the torsion tensor was first found in [\gates], in which a complete set of constraints for massless and massive IIA supergravity were
proposed. As far as we can tell, our results are in agreement with
those of [\gates], but it is not clear to us whether the $m$-dependence of the
4-form field strength was taken into account by these authors. 

When $m=0$ the IIA superspace constraints found above are just those obtained
by dimensional reduction of the standard D=11 superspace constraints. In
fact, they were deduced in this way in [\DHIS], independently of [\gates].
These constraints are known to imply the field equations of D=11 supergravity
[\brink]. Thus, the $m=0$ constraints imply the field equations of massless IIA
supergravity. It follows that the $m\ne0$ constraints imply the field equations
of the massive IIA theory. Note that by `constraints' we mean the specification
of the components of all superspace tensors of dimension 1/2 or less. The massive
IIA constraints are therefore formally identical to those of the massless
theory, differing only in the $m$-dependence of the $R\otimes R$ field strength
superforms. This is a consequence of the `natural' choice of basis of IIA
supergravity field variables selected by the coupling to the super D-2-brane.
We shall now conclude with a discussion of how this basis is related to the
`canonical' one, and why the new basis is simpler.

%%%%%%%%%%%%%%%%%%%%%%%%%%%%%%%%%%%

\chapter{Field variables in IIA/IIB supergravity} 

We first recall what the canonical variables are. To simplify the notation we 
use form notation and indicate the $NS\otimes NS$ 2-form by $B$ with
corresponding gauge transformation $\delta B = d\Lambda$. All other gauge 
fields are 
$R\otimes R$ potentials which we denote by $C^{(r)}\ (r=1,\cdots ,9)$. We use
the notation and conventions of [\BHO] but have renamed the
fields of IIA/IIB supergravity as follows:
$$
A^{(1)} \rightarrow C^{(1)}\, , \qquad
{\cal B}^{(2)} \rightarrow C^{(2)}\, , \qquad
C \rightarrow C^{(3)}\, ,\qquad 
D \rightarrow C^{(4)}\, .
\eqn\foura
$$
The potentials with $r \ge 5$ are the corresponding dual potentials.
The fields $C^{(r)}$ are
potentials of IIA (IIB) supergravity for $r$ odd ($r$ even).
In the canonical basis the $R\otimes R$ potentials transform under
the following $R\otimes R$ gauge transformations with parameters
$\Lambda^{(r)}\ (r=0,\cdots ,8)$ [\BHO]\foot{For simplicitly
we only give the rules for  $r=1,\cdots ,6$. The remaining $R\otimes R$
potentials can be dealt with similarly.}:
$$
\eqalign{
\delta C^{(1)} &= d\Lambda^{(0)} - {m\over 2}\Lambda\, ,\cr
\delta C^{(2)} &= d\Lambda^{(1)}\, ,\cr
\delta C^{(3)} &= d\Lambda^{(2)} + 2 d\Lambda^{(0)}B - m\Lambda B \, ,\cr
\delta C^{(4)} &= d\Lambda^{(3)} + {3\over 4} d\Lambda^{(1)}B
-{3\over 4} d\Lambda C^{(2)}\, ,\cr
\delta C^{(5)} &= d\Lambda^{(4)} - {15\over 2}d\Lambda^{(2)}B +
{15\over 2}d\Lambda C^{(3)}\, ,\cr
\delta C^{(6)} &= d\Lambda^{(5)} + \Lambda^{(3)}dB +\Lambda^{(1)} BdB
-{1\over 2} d\Lambda B C^{(2)} }
\eqn\fourb
$$
With respect to [\BHO] we have renamed the parameters as follows
$$
\Lambda^{(1)} \rightarrow \Lambda^{(0)}\, ,\qquad
\Sigma^{(2)} \rightarrow \Lambda^{(1)}\, ,\qquad
\chi \rightarrow \Lambda^{(2)}\, ,\qquad
\rho \rightarrow \Lambda^{(3)}\, .
\eqn\fourone
$$
The gauge transformations of the dual potentials $C^{(5)}$ and $C^{(6)}$ have
been taken from [\berone] and [\bertwo], respectively.

The new basis presented in [\ght] has the following distinguishing
features:

\noindent 1.\ \  None of the $R\otimes R$ potentials transform under the gauge
transformation of the $NS\otimes NS$ 2-form B (with parameter $\Lambda$)
except for the $m$-dependent terms in the IIA case.

\noindent 2. All $R\otimes R$ gauge transformations are written in a 
canonical way
such that in the terms containing the $NS\otimes NS$ 2-form B the
$R\otimes R$ parameter $\Lambda^{(r)}$ always occurs undifferentiated.

It is now straightforward to show that by performing a suitable redefinition
of the fields $C^{(r)}\ (r\ge 4)$ and the parameters $\Lambda^{(r)}\ 
(r \ge 2)$ the canonical basis \foura, \fourb\ can be transformed into the
new basis defined above. More precisely, the following 
redefinitions are needed:
$$
\eqalign{
C^{(4)\prime} &= C^{(4)} + {3\over 4} BC^{(2)}\, ,\cr
C^{(5)\prime} &= C^{(5)} - {15\over 2}BC^{(3)}\, ,\cr
C^{(6)\prime} &= C^{(6)} + {1\over 4} B^2 C^{(2)}\, ,}
\eqn\fourc
$$
and
$$
\eqalign{
\Lambda^{(2)\prime} &= \Lambda^{(2)} + 2B\Lambda^{(0)}\, ,\cr
\Lambda^{(3)\prime} &= \Lambda^{(3)} + {3\over 2}B\Lambda^{(1)}\, ,\cr
\Lambda^{(4)\prime} &= \Lambda^{(4)} - 15B\Lambda^{(2)} - 
15 B^2\Lambda^{(0)}\, ,\cr
\Lambda^{(5)\prime} &= \Lambda^{(5)} + {1\over 4}B^2\Lambda^{(1)}\, .}
\eqn\fourd
$$
In the new basis the $R\otimes R$ gauge transformations are given by
(omitting the primes)
$$
\eqalign{
\delta C^{(1)} &= d\Lambda^{(0)} - {m\over 2}\Lambda\, ,\cr
\delta C^{(2)} &= d\Lambda^{(1)}\, ,\cr
\delta C^{(3)} &= d\Lambda^{(2)} - 2 \Lambda^{(0)}dB - m\Lambda B \, ,\cr
\delta C^{(4)} &= d\Lambda^{(3)} + {3\over 2} \Lambda^{(1)}dB\, ,\cr
\delta C^{(5)} &= d\Lambda^{(4)} + 15\Lambda^{(2)}dB +
{15\over 2}m\Lambda B^2\, ,\cr
\delta C^{(6)} &= d\Lambda^{(5)} + \Lambda^{(3)}dB\, .
}
\eqn\foure
$$ 

As an example of the simplicity inherent to the new basis we will give the
T-duality rules of [\BHO] in this basis. First, to keep the calculations
simple,  we make the same assumption about the background fields as [\ght],
i.e.\foot{This assumption is not essential to the simplifications discussed
below.}
$$
g_{x\mu} = B_{x\mu} = 0\, .
\eqn\assumption
$$
Here $x$ refers to the isometry direction. Under this assumption
the $T$-duality rules of [\BHO] simplify as follows. The $T$-duality rules
for the $NS\otimes NS$ fields reduce to
$$
\eqalign{
{\tilde {g}}_{\mu\nu} &= g_{\mu\nu}\, ,\hskip 1.1truecm 
{\tilde {g}}_{xx} = 1/g_{xx}\, ,\cr
{\tilde {B}}_{\mu\nu} &= B_{\mu\nu}\, ,\hskip 1truecm
e^{2\tilde\phi} = e^{2\phi}/|g_{xx}|\, ,
}
\eqn\Buscher$$
while those of the $R\otimes R$ potentials are given by
$$
\eqalign{
{\tilde {C}}^{(0)} &= C_x^{(1)}\, ,\hskip 1.5truecm {\tilde {C}}_x^{(1)} 
= C^{(0)}\, , \cr
{\tilde {C}}_\mu^{(1)} &= -C_{x\mu}^{(2)}\, ,\hskip 1.2truecm 
{\tilde {C}}_{x\mu}^{(2)} = -C_\mu^{(1)}\, ,
\cr
{\tilde {C}}_{\mu\nu}^{(2)} &= {3\over 2}C_{\mu\nu x}^{(3)}\, ,\hskip 1truecm 
{\tilde {C}}_{\mu\nu x}^{(3)} = {2\over 3}C_{\mu\nu}^{(2)}\, ,\cr
{\tilde {C}}_{\mu\nu\rho}^{(3)} &= {8\over 3} C_{x\mu\nu\rho}^{(4)} - 
C_{x[\mu}^{(2)}B_{\nu\rho]}\, ,\cr
{\tilde {C}}^{(4)}_{x\mu\nu\rho} &= {3\over 8}\bigl [ C_{\mu\nu\rho}^{(3)} - 
C_{[\mu}^{(1)}B_{\nu\rho]}\bigr ]\, .
}
\eqn\BRR
$$ 
We see that under $T$-duality the $R\otimes R$ potentials $C^{(r)}$
transform  to the potentials $C^{(r\pm 1)}$
except for $C^{(3)}$ for which the T-duality rule involves the $NS\otimes NS$ 2-form $B$.
We find that in the new basis
all dependence on $B$ disappears. In particular
the rules involving $C^{(3)}, C^{(4)}$ are given by (omitting the primes)
$$
{\tilde {C}}_{\mu\nu\rho}^{(3)} = {8\over 3} C_{x\mu\nu\rho}^{(4)}\, ,
\hskip 1truecm
{\tilde {C}}_{x\mu\nu\rho}^{(4)} = {3\over 8} C_{\mu\nu\rho}^{(3)}\, .
\eqn\missing
$$

It is not too difficult to understand why, in the new basis, the
T-duality rules of the $R\otimes R$ potentials
are of the simple form given above. The point is that,
using an appropriate normalization, the kinetic term of any of the
$R\otimes R$ potentials takes the form (using the string-frame metric)
$$
{\cal L}_{R\otimes R} = \sqrt {|\hat g|}\, {(-1)^{p+1}\over 2(p+2)!}\,
{\hat R}^2_{p+2}(C)\, ,
\eqn\one
$$
where the hatted fields are ten-dimensional and $R(C)$ is defined as [\ght]
$$
R(C) = dC - dB \wedge C + me^B\, .
\eqn\two
$$
Because of the assumption \assumption, the reduction rules for the $R\otimes R$ 
potentials are particularly simple:
$$
\hat {C}_{\mu_1\cdots \mu_{p+1}} = {C}_{\mu_1\cdots \mu_{p+1}}\, ,\qquad
\hat {C}_{x\mu_1\cdots \mu_{p+1}} = {C}_{\mu_1\cdots \mu_p}\, .
\eqn\three
$$
Similar simple reduction rules apply to the curvatures.
Consider now  the kinetic  term for a IIA potential for fixed (even) p.
Reduction in the isometry direction $x$ leads to 
$$
{\cal L}_{\rm IIA} = 
\sqrt {| g|}\, {(-1)^{p+1}\over 2(p+2)!}e^{\chi/4}\,
{ R}^2_{p+2}(C) +
\sqrt {| g|}\, {(-1)^{p}\over 2(p+1)!}e^{-\chi/4}\,
{ R}^2_{p+1}(C)\, ,
\eqn\four
$$
with $\hat {g}_{xx} = -e^{\chi/2}$.
Similarly, reducing the kinetic term for a IIB potential for fixed
(odd) q leads to
$$
{\cal L}_{\rm IIB} = 
\sqrt {| g|}\, {(-1)^{q+1}\over 2(q+2)!}e^{-\chi/4}\,
{ R}^2_{q+2}(C) +
\sqrt {| g|}\, {(-1)^{q}\over 2(q+1)!}e^{\chi/4}\,
{ R}^2_{q+1}(C)\, .
\eqn\five
$$
with $\hat {g}_{xx} = -e^{-\chi/2}$.
Comparing these two expression for the two cases $q=p\pm 1$
immediately leads to the following simple T-duality rules for the
$R\otimes R$ potentials (together with the usual Buscher's rules for the
$NS\otimes NS$ fields)
$$
\eqalign{
\hat {C}_{\mu_1\cdots \mu_p} = \hat {C}_{x\mu_1\cdots \mu_p}\, ,
\hskip 1truecm
\hat {C}_{x\mu_1\cdots \mu_p} = \hat {C}_{\mu_1\cdots \mu_p}\, .
}
\eqn\six
$$
These are exactly the same T-duality rules as those given in [\ght].

\vskip 0.5cm
{\bf Acknowledgement}: We are pleased to thank Martin Cederwall, Bengt
Nilsson and Anders Westerberg for helpful discussions on issues related to the
work reported here.

\refout
\end